# A survey on Dependable Digital Systems using FPGAs: Current Methods and Challenges

Farah Natiq Kassab bashi

PG Scholar, College of Engineering, Department of Computer Engineering,
Mosul University, Iraq
Email: farah.enp84@student.uomosul.edu.iq

Shawkat Sabah Khairullah

Lecturer, College of Engineering, Department of Computer Engineering,
Mosul University, Iraq
Email: shawkat.sabah@uomosul.edu.iq

**Abstract:** *Fault tolerance is increasingly being use to design Dependable Digital Systems (DDS), which refers to the capability of a system to keep performing its intended functions in existence of faults. DDS are typically used in Safety-critical system (SCS) such as medical (I&C) devices, Nuclear power Plants (I&C) devices and Aerospace (I&C) systems, the failure in these systems can cause harm to environment, death, injury to people. Different fault tolerance techniques were developed to overcome these issues and that has led to increase the reliability and dependability of applications on Field Programmable Gate Arrays (FPGAs). In this paper, multiple related works are present dealing with different types of faults and fault tolerance methods in FPGA based systems. Furthermore, a comparison between the evaluation metrics of previous works of Fault Tolerant (FT) techniques like hardware redundancy overhead, time delay, reliability, and performance are also present.*

## 1. INTRODUCTION

SRAM field programmable gate arrays (FPGA) are integrated circuits (IC) that can perform any digital logic task. An FPGA formation of block memories (BRAM), flip-flops (FF), lookup tables (LUT), other specific resources like (MGTs, DSPs, etc.) and a big configurable routing network to link all these components programmatically [19]. FPGAs applied in high radiation environments, like high-energy physics environments or space, SRAM FPGAs, are sensible to radiation, ionizing and can suffer of single-event upsets (SEUs) in the FPGAs internal state, inclusively the configuration RAM (CRAM), the CRAM bits dominance the operation of the device, like routing information or the contents of LUT. An accuracy of SEU in one of these bits have the ability to change the system's action. When the suggest design next employ this bit, a faulty value will increment out of the LUT causing an incorrect computation or circuit failure [16].

SRAM-based in (FPGAs) are utilized widely in real life implementations, such as in space, autonomous driving, and high-tech systems, where high dependability is an involuntary need. We know that the design of FPGA is stocked in the CM (Configuration Memory) in SRAM-based FPGAs, thus they are sensible to SEUs (Single Event Upsets), when the latest happen, a memory cell is overthrown from a logical '1' to a logical '0', or from a '0' to a '1', due to ecological factors. Furthermore, FPGAs are expose to Multiple Event Upsets (MEUs) beside (SEUs) because of continual shrinking of the dimensions in the transistor [20], so it is very significant to make the FPGAs design more Fault Tolerant.

Safety-Critical digital systems must be design to be reliable against different types of failure modes. The reliability of these systems is evaluated using Markov modeling chains to ensure that the design specifications are met [22]. The behavior of a Safety-Critical digital system can be modeled using two states (normal state and degraded state) such as it shown in *Figure 1*. We can say that the system is in normal state when no faults appear; if the latest happen, it will be in one of two ways (fail-safe state and fail unsafe state). Faults that happen in safe state are those faults that can be Detected and Tolerated, on the other word have System Recovery (SR) in opposite of Faults that happen in Unsafe state (Catastrophic System Event (CSE)) that cause Serial injury to people, Death or Damage to the environment and could not be Detected and Tolerated. Faults can classify according to the duration of the fault into: Permanent faults (PF), Transient faults (TF), Intermittent faults (IF) and common cause faults (CCF) [21].

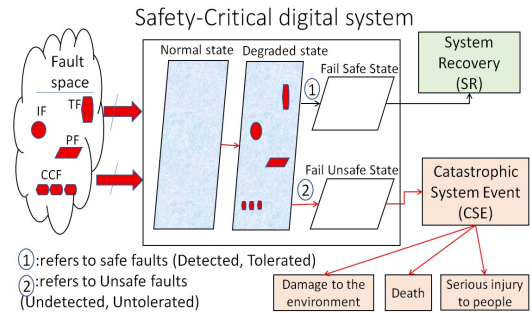

*Figure 1 Safety-critical digital system*

Understanding the correct behavior of a digital embedded system led to developing effective fault-tolerant techniques against different fault types [22]. This article studies the fault tolerant mechanisms and their effectiveness on the performance, area overhead, power consumption and time delay of the system to obtain Dependable Digital Systems (DDS) for safety-critical computer systems leverage existing Field Programmable Gate Arrays (FPGAs) technology.

Section 2 explains the classification of hardware faults. In section 3, popular fault tolerance methods are defined. In section 4 and section 5, the mechanisms of hardware redundancy and an overview of current fault tolerance methods is present. Finally, the design challenges, conclusion and future works are presented in section 6 and section 7.

## 2. CLASSIFICATION OF HARDWARE FAULTS

We can define a Fault that is a weakness in the hardware or software in the system that cause an incorrect state. When an error affect one component or more of the System, it is possible that the same error will spread and cause other errors; furthermore, that error may spread to service interface and deviates the function of the system causing system failure.

In space, the environment Radiation cause many faults. Faults can be classified as permanent and temporary, Temporary faults can be intermittent or transient [15].

### 2.1 Transient faults:
It is present for a short time and then hide.

### 2.2 Intermittent faults:

Once it is present then it is hide and then it is reappeared.

### 2.3 Permanent faults:
When it is present, it remains present. In TABLE I we can see the main difference between the three types of faults.

TABLE I [15]

| Fault type | The fault cause | Notes |
|---|---|---|
| Transient faults or soft error | Environmental conditions:<br>-electromagnetic interference<br>-injection of alpha and neutrons particles<br>- power supply<br>-electrostatic discharge<br>-interconnect noises | They do not led to any permanent damage |
| Intermittent faults | the presence of any marginal hardware:<br>-higher temperature<br>-higher voltage | If these type of faults continue for a long time that will cause permanent faults. |
| Permanent faults | some hardware physical defects such as:<br>-shorts in a circuit<br>-broken interconnections<br>-stuck cells in a memory. | These faults are effective during the device operation and cause the aging of it. |

## 3. POPULAR FAULT TOLERANCE METHODS

We can define fault tolerance that it is the capability of a system to keep performing its designed functions in existence of faults. In other word, we can say that fault tolerance is connect with reliability, successful operation, and absence of malfunctions.

There are different approaches to realizing fault tolerance such as fault detection, faulty component diagnosis and isolation, and modular redundancy [18]. The most popular method is specific quantities of redundancy; two types of redundancy are potential [6]:

### 3.1 Space redundancy:
Supply further components, data items, or functions that are not necessary for fault-free procedure. Space redundancy can be classified according to the redundant resources in the system that added as software, hardware, and information redundancy.

### 3.2 Time redundancy:

Hither the data transmission or computation is reduplicate then the outcome is compared with a stoked copy of the previous outcome.

## 4. HARDWARE REDUNDANCY MECHANISMS

Hardware redundancy can be achieved by supply two physical copies or more of the hardware component, and it might be the only method to progress the systems dependability. Hardware redundancy can be classified as active, passive, and hybrid.

### 4.1 Passive redundancy:

In this approach faults are masked rather than they are detect, insuring that in spite of a fault presence only the values that are correct will pass to the output of the system. As an example, we take Triple Modular Redundancy (TMR) shown in *Figure 2*, which consist of three copies of the FPGA slices, synchronous timing clock and a voting circuit to mask the errors.

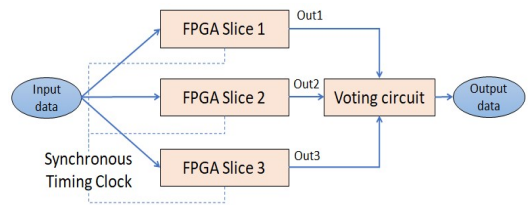

*Figure 2 Triple modular redundancy.*

We can use TMR in different ways in the design. Some applications use TMR as partial application in critical parts of the design whose failure can cause systems-wide failure. Other applications might require a full module of TMR implementation that satisfy special reliability requirements [13].

### 4.2 Active redundancy:

It achieves a fault tolerance through detecting the faults that occur firstly then it will perform an action that is need returning the system to a normal operation state, active redundancy techniques are: standby, duplication with comparison, and pair-and-a-spare.

### 4.3 Hybrid redundancy:

It merges between active and passive approaches. We can use fault masking to block obstetric of erroneous results. Fault (detection, location, and recovery) are employed to exchange the faulty component through a spare. Hybrid redundancy can enable reconfiguration with no downtime of the system and usually they use in safety-critical applications.

In TABLE II, we can see examples about the three types of redundancy and their usage.

## 5. OVERVIEW OF CURRENT FAULT TOLERANCE METHODS

Field-programmable gate arrays based on static-random access memory (SRAM), we can reprogram it one after other as necessary that make it more flexible and widely usable in different applications. However, such flexibility depends on SRAM and its cells programming, that are sensitive to different perturbations; such as Single-Event-Upsets (SEUs) and Multiple-Event-Upsets (MEUs), so different FPGA hardening techniques were proposed, in TABLE III we can see the main difference between them.

TABLE II HARDWARE REDUNDANCY TYPES [6]

| Type of Hardware redundancy | Their usage | Examples |
|---|---|---|
| Passive redundancy | In high-reliability applications | -aircraft flight control systems<br>- embedded medical devices (heart pacemakers and deep-space electronics) |
| Active redundancy | In high Availability applications | -time-shared computing systems<br>-transaction processing systems |
| Hybrid redundancy | In safety-critical applications | - control systems for chemical processes<br>- nuclear power plants<br>- weapons<br>- medical Equipment<br>- aircrafts<br>- trains<br>- automobiles |

An Inspired master-slave TMR technique for fault-tolerance of SRAM-based FPGA was proposed by Lahrach, et al. (2010) [1], this method combined between master-slave technique and triple modular redundancy and take advantages of partial reconfiguration to make tolerate to permanent faults. A novel technique for detection and diagnosis of FPGAs permanent faults, on-line built-in self-test (BIST) have proposed by Meshram et al. (2011) [2].

In this method, the arrangement of bit-stream can modified by hardware controller existing on the same chip, they use redundant device and a spare to replace faulty device. This technique has low fault latency, very high fault coverage and makes support to the theoretical analysis, although maximum (100%) yield can get from this method, using spare device will increase the area of the chip and effect the time delay. Luo et al. (2011) [3], have proposed a payload system that can detect and correct Single-Event-Upset (SEU) for SRAM-based FPGA, the system can detect internal SEU via partial reconfiguration and scrub to avert serious impact. This method will not take inner re source of the FPGA destination and not affect the systems normal operation, furthermore the system demand less time and the scrub frequency can change according to stability requirements. Ashraf et al. (2011) [4], have used different designs source of the FPGA destination and not affect the systems normal operation, furthermore the system demand less time and the scrub frequency can change according to stability requirements. of TMR to evaluate its reliability improvement in comparable time cost and area using design variety for common-mode-faults (CMF) and Random-single-faults (RSF). Namely, multiple implementations were evaluated of the same design functionality by using repository of methods like: Case-Based, Templates, NAND/NOR-Based and Inverted-Output methods. The usage of one of these designs or combination of them will increase the TMR-based system's fault-tolerance and show more reliability to CMF exposure without more run-time cost or efforts, but there is a challenge when transaction with sequential circuits because the product does not depend only on the inputs.

TABLE III THE MAIN DIFFERENCE BETWEEN THE STUDIES

| Study No. | FPGA board | Hardware overhead | Time delay | Reliability | Performance |
|---|---|---|---|---|---|
| [1] | N/A | Low overhead amount of CLBs not exceed 2.7% | Management the runtime | Improve the reliability | Complete availability and detect single and double faults |
| [2] | FPGA Altera Quartus II EP2S158484C3. | Increase the area of the chip | Effect time delay | Improve the reliability | Very high fault coverage (100%) yield |
| [3] | Xilinx Virtex-4 FPGA | not take inner resource of the FPGA | Less time | Improve the systems Reliability | Improve the systems stability. |
| [4] | Xilinx Virtex-4 FPGA Device | Less area | Less run-time cost | More reliability to CMF | Increase the fault-tolerance |
| [5] | N/A | Low overhead | N/A | N/A | Focus on the execution-time |
| [7] | The Virtex-5 XC5VLX50T FPGA | N/A | N/A | N/A | low power overhead |
| [8] | FPGA Virtex5-XC5VSX50T on ML 506 board | Limited Redundant Area | N/A | N/A | Mitigate transient fault and permanent fault |
| [9] | MCNC'91 benchmarks | 100% hardware overhead | Shorter runtime | reduce 90% SEU faults | It use for detection only Very high SEU immunity level at reasonable area overhead |
| [10] | MCNC'91 benchmarks | 100% hardware overhead | N/A | Reduce 70% of SEU faults Not require high reliability in specific area | It use for detection only |
| [11] | Artix7 XC7A100T FPGA. | Can be save up to 15% of the resources | N/A | Improve Reliability but not supply full coverage | Improve the stability |
| [12] | Altera EP4CE115F29C7 FPGA device | Occupy less area. | N/A | The reliability Increase | Highest performance |
| [13] | Virtex device. | 200% hardware overhead | Faster recover (too much) | Reliability is improve compared with simplex voter | Faster recover for itself compared with a scrubbing |
| [14] | Microsemi ProASIC3E series Flash based FPGA | Low overhead. | N/A | Enhance the systems reliability | Effective in treatment various faults. |
| [17] | XC5VLX110T Virtex-5 FPGA | 200% hardware overhead | Recover quickly | Improve the reliability if it compared with simplex voter | Improve the fault-tolerance if it compared with simplex voter Can tolerate single fault and multiple faults |
| [20] | Xilinx 7 series FPGAs | Increase 20% extra area after half year | N/A | The reliability 18% Increase after half year | Increase the performance at the expense of hardware overhead |
| [23] | XC7Z020 FPGA. | Less occupied area of FPGA device | N/A | More reliable than NMR | Improve the systems stability. |

Kshirsagar et al. (2012) [5], have presented a novel FT technique in FPGA-based systems, horse-shifting allocation. This technique can done by only shifting vertically or horizontally without effecting the physical design of the application and without load other ordered data from FPGAs on-chip. On the other hand, some CLBs are confine as spare and they use to replace with the detective CLBs, the spare blocks depend on the number of CLB. In this study they deal with faults happened by degradation, manufacturing defects, Single-Event-Upsets (SEUs) and Single – Event-Transients (SETs), This FT technique focus on the execution-time that is 14672ms.

Tarrillo *et al.* (2013) [7], have proposed an N-modular redundancy fault-masking capability and synthesized it into SRAM-based FPGA, different redundant modules were used ranging from three copies in (standard TMR) up to seven copies. In this method a lot of redundant copies were used, that lead to reduce the scrubbing rate significantly with a low power overhead; as a result, will reduce the power consumption of overall FPGA, in spite of power consumption not increases lineal proportion with the no. of redundant modules in nMR.

Miculka et al. (2013) [8], have present a fault tolerant method in FPGA that have the ability to mitigate transient fault and sundry permanent fault that occur sequentially. To mitigate transient fault, they use partial-dynamic-reconfiguration (PDR) and for permanent faults mitigation they use a specific FT architecture that occupy less resources than the already used and exclude the FPGAs faulty part. In this technique, they based on the stored precompiled configurations in an external memory and they used the relocation technique to reduce the space that wonted for a partial bit stream. This method can handle permanent faults which is affect two separate Partial Reconfigurable Modules PRMs, if greater than two PRMs are affected except PRM (ROUTE), the fault tolerant system cannot produce correct outputs. For longer availability, fault tolerant system with more than five PRMs can be design.

A new type of Dual-Module-Redundancy (DAO) design in FPGA was proposed by Zheng et al. (2015) [9], in this design the duplication will operate in LUT level then a logic (OR/AND) voter will vote each pair of the identical LUTs. The LUT duplication will detect the transfer of faults to the following hardened level of the circuit furthermore; this architecture does not need to reset or add circuits for error judgment to switch to un fault part, the proposed algorithm has much shorter runtime than other methods.

A Fault Masking Dual Module Redundancy (FMDMR) structure for FPGA was proposed by Zheng et al. (2016) [10]. In this method they use (AND/OR) logic as dual-module-redundancy voter and add it after LUT pair duplication, the voter's insertion will not cause additional hardware overhead but this method can use for fault detection only when the fault is explored, behind it cannot offer auto switch fault localization or fault-free module.

The method proposed in [9] is similar to the work proposed by Zheng et al. (2016) [10], both methods have 100% hardware overhead, and they use for fault detection only, but this method have an improvement of reduction SEU faults 20% over the method proposed in [10] and have very high SEU immunity level at a reasonable area overhead.

Sanchez-Clemente et al. (2016) [11], have propose a new way to build a Partial TMR for FPGAs by the use of approximate logic Circuits to minimize the resource utilization. This method have the ability to provide a selective protection against unidirectional errors, furthermore; it is scalable and have a flexibility to balance between the reliability and the overheads beside cost reduction, this method can provide more safeguard than full TMR if the errors are not correct and may accumulate.

Mallavarapu et al. (2017) [12], have proposed number of configurations of 5MR such as conventional 5MR, 5MR with XNOR, 5MR with XOR, 5MR using cascaded TMR, 5MR with (4 to 1) multiplexer. 5- MR with XOR occupies the lowest logic elements with highest performance than others; the 5-MR with XNOR consumes lower power if it compared with the others.

Afzaal et al. (2017) [13], have proposed a Self-checking LUT based Voter in TMR to overcome the failures that can happened in presence of reduction voter, multiple redundancies attach on the voter in a single output line to overcome these issues. In a proposed voter, there is an improvement in the reliability if it compared with simplex voter, which lead to much faster recover for itself compared with a scrubbing for the same, on the other hand the drawback of this method is the Arbiter switching functionality which can be influenced by an upset.

Unni et al. (2018) [14], have present an efficient configurable watchdog timer (WDT) and its realization in FPGA, this watchdog timer can be use in safety-critical applications. The proposed watchdog hardware can interface to different systems and processors with minor modifications, fault detection mechanisms can build into the watchdog and adds robustness to it. When the system enters in fault state and continually resets the timer, here the error state will not be detected. Furthermore, hardware failures appear when the watchdog timer did not get service from the processor or due to a mistake in the processor service because of interrupt overloading, transient faults or intermittent failures.

Afzaal et al. (2018) [17], have proposed a self-checking TMR voter for reliable on-chip assent voting that can make a toggle between multiple redundancies of the voter that attached the same output and thus which they recover quickly from an upset, this work is an extension of [13]. This method can tolerate single fault, and tolerate multiple faults if they not lead to overlapping errors, furthermore this method improves the fault-tolerance and reliability if it compared with simplex voter, but the same drawback as [13] appear in the functionality switching of the arbiter.

The method proposed in [1] has three benefits if it compared with redundancy-based fault-tolerance present in [13] and [17]: the option of management the runtime, complete availability and the ability to detect single and double faults with low overhead that lead to increase the reliability at the expense of architecture complexity.

If we compare the work presented in [11] with [13] and [17] we can see that there is a main drawback of this method, because it does not supply full coverage contra single faults if it compared with conventional TMR as present in [13] and [17].

A Penta and Hexa Modular Redundancy for FT technique was proposed by Shaker et al. (2019) [20], Penta Modular Redundancy (5MR) can recover from SEUs beside MEUs but at the same time there are some faulty scenarios that cannot be detected, so they propose Hexa Modular Redundancy (6MR) FT technique. This method increases the reliability of system but the cost of extra-added hardware will increase and the hardware overhead will increase.

If we compare the proposed work in [7] with [20] and [12], we can see that this method reduces the power consumption of overall FPGA, while [20] and [12] increase the reliability of system.

Farias et al. (2020) [23], have propose a resilient hardware architecture that can use in critical systems, which is TMR-with-spares, an active redundancy model in FPGA, this architecture, can port to different models in FPGA without compromise the performance, availability, and reliability. In this architecture the developed controller, give it the possibility to adjust between TMR with one spare or two spares. This method is more reliable than NMR present in [20] and [7], furthermore; the proposed architecture can respond appropriately even three failed modules happened. This method is able to storage the information of a faulty module and returning to previous configuration while these transient faults disappear.

## 6. CHALLENGES

The current state of the practices for realizing dependable digital systems is achieve using different modular fault tolerance mechanisms such as triple modular redundancy, duplication with comparison, and hybrid redundancy. These traditional mechanisms have worked well but have led to some design challenges, as they are list below.

### 6.1 Area overhead:

Most of the fault tolerance techniques are depend on repetitions, either by adding more copies that are redundant or by adding spare cells that help to make recover of the faulty blocks. To increase the reliability in modular redundancy it is required to add more hardware components, in other word as the number of modular redundancies increase the reliability will increase but at the expense of increasing the area overhead. In other designs, they combine between modular redundancy and number of spare cells; this method can provide more reliability than NMR and reduce the area overhead. Therefore, the challenge is to obtain maximum fault-tolerance with a minimum area overhead.

### 6.2 Power consumption:

The power consumption depends on fault tolerance technique that have used. As the active replication number increases, the power consumption will increase. As an example, TMR techniques have one-third greater power than DMR. The challenge of fault tolerant is to recover the fault in the system with higher reliability and lower power consumption.

### 6.3 Time delay:

Time delay depend on the fault tolerance method that have chosen. Therefore, an optimization should be happened about this method in terms of time consumption to make faster fault tolerance system with lower time delay.

## 7. CONCLUSION AND FUTURE WORKS

In this article, a survey on several research works have been presented to design fault-tolerant hardware strategies, mainly to address the current fault tolerance methods and challenges. Some studies have presented methods to mitigate faults in hardware using modular redundancy like DMR, TMR and NMR. Some researchers used DMR as mitigation technique, but this method can be used for fault detection only and cannot recover from faults, other researchers used Partial TMR, but this method does not supply full coverage. Furthermore, NMR FT technique has led to increasing the reliability of the system at the expense of extra-added hardware overhead. To increase reliability with less hardware overhead led researchers to investigating for other types of FT techniques that combine between modular redundancy and active redundancy, like TMR-with-spares. On some occasions, these methods will provide us the same reliability as NMR or more at the expense of architecture complexity and area overhead.

Future research works should focus on different topical areas which include the verification and validation (V&V) of safety-critical properties for the digital (I&C) application using FPGA verification tools. In addition, the usage of modular hybrid redundancy with active spares to tolerate transient, intermittent, permanent faults, and common cause failures must be investigated. Finally, optimizing the design of fault tolerant system by reducing the number of spares as possible to increase the reliability and performance and reduce the hardware area overhead is necessary in dependable computing.

**Authors Biography**

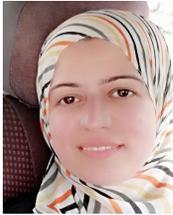

**Farah Natiq Kassab bashi,** received the B.S. degree in Computer Engineering from University of Mosul, college of engineering, Mosul, Iraq in 2004. Currently, she is working for M.Sc. Degree in Computer Engineering at Mosul's University College of Engineering. She has employed at the University of Mosul, College of Agriculture and Forestry in Iraq since 2007 until now, her research interests include fault-tolerant system design, VLSI, and FPGA-based systems.

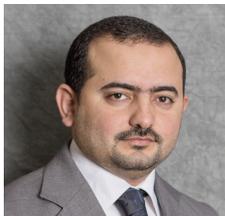

**Shawkat Sabah Khairullah,** received the B.S. and M.Sc. degrees in computer engineering from the University of Mosul, college of engineering, Mosul, Iraq in 2006 and 2011. He received the Ph.D. degree in Computer Engineering from Virginia Commonwealth University, Richmond, USA in 2018. He is a Lecturer at the University of Mosul, Iraq. Dr. Shawkat research interests include dependable digital system design, VLSI, bio-inspired self-healing hardware systems, and FPGA-based systems.